\documentclass[aps,prl,twocolumn,superscriptaddress,showpacs]{revtex4}

\usepackage{graphicx}

\begin{document}

\title{Comment on ''Superfluid turbulence from quantum Kelvin wave to classical 
Kolmogorov cascade''}

\author{Giorgio Krstulovic}
\affiliation{Laboratoire de Physique Statistique de l'Ecole Normale 
Sup{\'e}rieure, \\
associ{\'e} au CNRS et aux Universit{\'e}s Paris VI et VII,
24 Rue Lhomond, 75231 Paris, France}
\author{Marc Brachet}
\affiliation{Laboratoire de Physique Statistique de l'Ecole Normale 
Sup{\'e}rieure, \\
associ{\'e} au CNRS et aux Universit{\'e}s Paris VI et VII,
24 Rue Lhomond, 75231 Paris, France}
\date{\today}


\maketitle
%\section*{Intro}
In a recent Letter \cite{yepez:084501}, Yepez {\it et al.}
performed numerical simulations of the Gross-Pitaevskii equation (GPE) using a novel unitary quantum algorithm with very high resolution. 
They claim to have found new power-law scalings for the incompressible kinetic energy spectrum:
"...(the) solution clearly exhibits three power law regions for $E_{\rm kin}^{\rm incomp}(k)$: for small $k$ the Kolmogorov $k^{-\frac{5}{3}}$ spectrum while for high $k$ a Kevlin wave spectrum of $k^{-3}$...". 

In this comment we point out that the high wavenumber $k^{-3}$ power-law observed by Yepez {\it et al.} is an artifact stemming from the
definition of the kinetic energy spectra and is thus not directly related to a Kelvin wave cascade.
Furthermore, we clarify a confusion about the wavenumber intervals on which Kolmogorov and Kelvin wave cascades are expected to take place.
Finally, we point out that the incompressible kinetic energy spectrum of the initial data chosen by Yepez {\it et al.} scales like $k^{-1}$ at small wavenumbers, perhaps not the best choice to obtain a clean Kolmogorov regime.

%\section*{Basic definitions}
The dynamics of a superflow is described by the GPE
\begin{equation}
\partial_t{\psi}= i c/(\sqrt{2} \xi)(\psi  -|\psi|^{2} \psi 
+ \xi^2 \nabla^2 \psi ), \label{eq:esnl}
\end{equation}
where the complex field $\psi$ is related by
Madelung's transformation
$\psi= \sqrt{\rho} \exp \left(i \frac{\phi}{\sqrt{2} c \xi} \right)$
to the density $\rho$ and velocity $\vec{v}= \nabla{\phi}$
of the superfluid.
In these formulae, $\xi$ is the
coherence length and $c$ is the
velocity of sound (for a fluid of 
unit mean density).
The superflow is irrotational, except 
on the nodal lines 
$\psi=0$ which are the superfluid vortices. 

The GPE dynamics Eq. (\ref{eq:esnl}) conserves the energy that can be written as the sum (the space-integral) of three parts:
the kinetic energy
${\cal E}_{kin} = 1/2 (\sqrt \rho v_j)^2$, the internal energy 
${\cal E}_{int}=(c^2/2) (\rho -1)^2$
and the quantum energy ${\cal E}_{q}= c^2 \xi^2 (\partial_j \sqrt{\rho} )^2$.
Using Parseval's theorem, one can define corresponding energy spectra:
e.g. the kinetic energy spectrum
$E_{kin}(k)$ as the angle-average of
$\left|\frac{1}{(2\pi)^3} \int d^3 r e^{i r_j  k_j}\sqrt \rho v_j \right|^2$ \cite{nore:3896}.

The 3D angle-averaged spectrum 
of a smooth isolated vortex line is known to be proportional
to that of the 2D axisymmetric vortex, an exact solution of Eq. (\ref{eq:esnl})
given by $\psi^{vort}(r)= \sqrt{\rho(r)}
\exp(\pm i  \varphi)$ in polar coordinates
$(r,\varphi)$. The corresponding velocity field $v(r)= \sqrt{2} c \xi/r$ is azimuthal and 
the density profile, of characteristic spatial extent $\xi$, verifies $\sqrt{\rho(r)}
\sim r$ as $r \to 0$
and $\sqrt{\rho(r)} = 1+ O(r^{-2})$ for $r \to \infty$.
Thus  $\sqrt \rho v_j$ has a small $r$ singular behavior of the type
$r^0$ and behaves as $r^{-1}$ at large $r$. In general for a function scaling as $g(r) \sim r^{s}$ 
the (2D) Fourier transform is $\hat{g} (k) \sim k^{-s-2}$ and the associated spectrum scales as $k^{-2 s -3}$.
Thus $E_{kin}(k)$ scales as $k^{-3}$ for $k \gg k_{\xi}\sim \xi^{-1}$ and as $k^{-1}$ for $k \ll k_{\xi}$. \cite{nore:2644}.

%\section*{$k^{-3}$-scaling}
Following the above discussion, the $k^{-3}$ power-law observed in \cite{yepez:084501} is an artifact stemming from the
definition of the kinetic energy spectra and is not directly related to a Kelvin wave cascade.

%\section*{wavenumber scales associated with the different cascades}
Another very important scale, not discussed in the Letter \cite{yepez:084501},
is the scale $\ell$ of the mean intervortex distance.
The energy cascade is expected to end at $k_{\ell} \sim \ell^{-1}$ \cite{nore:3896} and the Kelvin wave cascade to 
begin there, after an eventual bottleneck \cite{lvov:024520}. Note that $ \ell_I \gg \ell \gg \xi$, 
where $\ell_I$ is the energy containing scale.
We thus believe that nothing particularly interesting is taking place
between $k_\xi$ and the maximum wavenumber $k_{\rm max}$ of the simulation and that there is
a confusion in  \cite{yepez:084501} between $k_\ell$ and $k_\xi$.

%\section*{Compressible Energy} 
Furthermore, the initial data used in \cite{yepez:084501} (see the supplementary material)
is a 3D set of 12 straight vortex lines,
with intervortex distance $\ell$ of the order of the box size.
The $k^{-1}$ scaling of the initial data $E_{kin}(k)$ thus extends down to small wavenumbers $k_I \sim \ell_I^{-1}$.
This behavior of the initial data is in contrast to the Taylor-Green initial data used in \cite{nore:2644}, where destructive interferences
deplete the value of $E_{kin}(k)$ in the interval $k_I<k<k_\ell$. 
The initial data chosen in \cite{yepez:084501} is thus perhaps not the best choice to obtain a clean Kolmogorov regime.
This might explain the high
level of compressible kinetic energy in the $-5/3$ scaling zone
that is apparent in the movie of the supplementary material.

We thank M. Abid and C. Nore for useful discussions.

%\bibliographystyle{unsrt}
%\bibliography{bibli}

\end{document}